# The transmission of liquidity shocks via China's segmented money market: evidence from recent market events[†]

Ruoxi Lu[*], David A. Bessler, and David J. Leatham

Department of Agricultural Economics, Texas A&M University, 2124 TAMU, College Station, TX 77843-2124, USA

This version: May 18, 2018

**Abstract**

This is the first study to explore the transmission paths for liquidity shocks in China's segmented money market. We examine how money market transactions create such pathways between China's closely-guarded banking sector and the rest of its financial system, and empirically capture the transmission of liquidity shocks through these pathways during two recent market events. We find strong indications that money market transactions allow liquidity shocks to circumvent certain regulatory restrictions and financial market segmentation in China. Our findings suggest that a widespread illiquidity contagion facilitated by money market transactions can happen in China and new policy measures are needed to prevent such contagion.

*JEL Classification:* C32; E44; E58; G01; G12; G2
*Key words:* Money market; Interbank market; Repo; Shibor; Liquidity shock; Transmission

---

[†] https://doi.org/10.1016/j.intfin.2018.07.005
[*] Corresponding author.
  E-mail addresses: ruoxi.lu@gmail.com (R. Lu)

# 1. Introduction

The interconnectivity within China's financial system is generally perceived to be limited compared to the financial systems in the U.S. and the other advanced economies. The direct interactions among different types of financial market participants across different asset markets in China can sometimes be highly restricted or even impossible, due to various degrees of market segmentation and regulatory controls. Financial institutions in China, particularly the banking sector[1], are heavily regulated with regards to which market they can access, who they can trade with and who they can finance. In such a system, it is seemingly difficult for a liquidity shock to transmit between two different parts of the system that are not directly connected. However, the perception may be different if we take into account the rapid expansion of China's money market in recent years and the potential transmission paths for shocks created by money market transactions.

Brunnermeier (2009) and Gorton and Metrick (2012) highlight the critical role of money market in transmitting liquidity shocks during the 2007 U.S. subprime crisis. Although China has yet to experience a system-wide liquidity crunch of similar scale, the risk of a widespread illiquidity contagion in China may have been growing along with its booming money market. We are particularly concerned about China's closely-guarded banking sector. The money market transactions involving banks and other investors, such as repurchase agreements (repo), can create transmission paths for liquidity shocks to spread between the banking sector and the rest of the financial system, despite the market segmentation and regulatory restrictions that partially isolate the banking sector. The existence of such pathways could pose a substantial threat to the overall stability of China's financial system, but it has not been assessed by the previous finance literature on China.

The primary objective of this paper is to identify the transmission paths created by money market transactions in China that are capable of spreading liquidity shocks into and out of China's banking sector, and to empirically assess how these pathways have functioned over the last five years. Specifically, we want to examine their involvement in the two dramatic financial market events that took place in China during this period, namely the rollercoaster ride in China's stock market between late-2014 and mid-2015 and the "Shibor[2] Shock"[3], a short-lived but severe cash crunch in China's banking sector in mid-2013. Given the fact that China's banking sector has always been strictly prohibited from trading in the stock market, these two events provide an ideal opportunity to explore whether or not a liquidity shock originated from a market inaccessible by the banks could affect the liquidity condition of banking sector via the money market, and whether or not a liquidity shock originated from the banking sector could affect the liquidity condition of a market that the banks do not have direct access to via the money market.

---

[1] The banking sector in China does not include investment banks, which are called securities firms in China.

## 2. Background

Before presenting the empirically analysis, we want to first provide a theoretical background on how money market transactions transmit liquidity shocks across financial assets and financial institutions, and how the transmission paths may function differently in China due to the unique characteristics of China's financial system. To avoid ambiguity, we explicitly specify the realm of money market in this paper as the wholesale credit market for short-term debt securities and credit instruments with maturities of one year or less.

### 2.1. The transmission of liquidity shocks via money market transactions

According to Brunnermeier and Pedersen (2009), the concept of liquidity can be divided into two different aspects: the market liquidity of assets, i.e. the ease with which they are traded, and the funding liquidity of traders, i.e. the ease with which they can obtain funding. The interactions among the market liquidity of multiple financial assets and the funding liquidity of multiple traders enable the transmission of liquidity shocks far beyond their origins. The money market is a major venue where the two types of liquidity intertwine with each other. On one hand, the money market serves as the main platform for the institutional trading of highly liquid and relatively safe short-term credit assets, such as treasury bills and commercial papers. On the other hand, the money market provide a medium for institutional investors to borrow short-term funds from each other with collateralized and uncollateralized instruments, such as repos and call loans, in order to finance new trades or support the crucial functions in their day-to-day operations.

---

[2] Shibor is abbreviated from Shanghai Interbank Offer Rate, the Chinese equivalent to Libor.
[3] This name of the event was first used in The Economist (2013a).

Based on the two aspects of liquidity, we characterize the transmission paths of liquidity shocks facilitated by money market transactions into three types: 1. the co-movement of funding illiquidity between lenders and borrowers; 2. the co-movement between the market illiquidity of assets and the funding illiquidity of their traders; and 3. the co-movement of market illiquidity across multiple assets. According to Brunnermeier and Pederson's (2009) theoretical model, these illiquidity co-movements will strengthen when a liquidity shock spreads among assets and traders, and will weaken once their market/funding liquidity improves.

We use a hypothetical example to illustrate how the three transmission paths function in the event of a liquidity shock. Brunnermeier and Pederson (2009) suggest that the increasing market illiquidity of an asset will lead to its increasing price volatility, which in turn will drive up the repo haircut[4] or margin requirement for the asset to be used as collateral. Suppose a financial institution invests in an asset and uses it as a repo (or margin) collateral to borrow short-term funds and make new trades. When the asset used as collateral is affected by a liquidity shock, its repo haircut (or margin requirement) rises, forcing the institution to raise new funds from the money market and withdraw some of its financing to other traders, subsequently reducing the funds available for other traders.

---

[4] Repo haircut is the difference between the market value of repo collateral and the amount of funds obtained by repo

When the funding liquidity of these traders worsen to an extent that they can no longer borrow from the money market, they are then forced to fire-sale their assets, in turn driving up the market illiquidity of multiple assets simultaneously. If these affected assets are also used as repo (or margin) collaterals by other traders, then the liquidity shock will be transmitted to these traders and the assets in their portfolio as well, ultimately resulting in a system-wide illiquidity contagion. In fact, if we replace the initially affected asset and the traders in this hypothetical example by the asset-backed commercial papers (ABCPs) tied to subprime mortgages and Wall Street banks, respectively, our example becomes a partial illustration of how liquidity shock was spread across the U.S. financial system during the 2007 subprime crisis. This is documented with more details in Brunnermeier (2009) and Gorton and Metrick (2012).

**2.2. The obstructed transmission paths in China's segmented money market**

According to the framework specified in Section 2.1, if we ignore the restrictions and segmentation in China's financial system, there are three potential pathways through which money market transactions can transmit liquidity shocks between the banking sector and the stock market in China: 1. through the interaction between the funding illiquidity of the banking sector and the funding illiquidity of stock traders, if the banks provide financing to stock traders via the money market; 2. through the interaction between the funding illiquidity of the banking sector and the market illiquidity of equity assets, if the banks trade stocks with funds borrowed from the money market; and 3. through the interaction between the market illiquidity of stocks and the market illiquidity of assets used by banks as collateral to borrow from the money market. However, in reality, the second path is nonexistent because the banks in China are barred from trading in the stock market, and the third path is severely obstructed by the segmentation in China's money market, leaving the first path the most susceptible for transmitting liquidity shocks between the banking sector and the stock market. In the rest of this subsection, we explain in detail why this is the case and discuss how China's money market structure affects how we conduct our analysis.

Based on China's money market structure, we visualize how China's banking sector interacts with other traders via financing and trading activities in China's segmented money market with Figure 1.

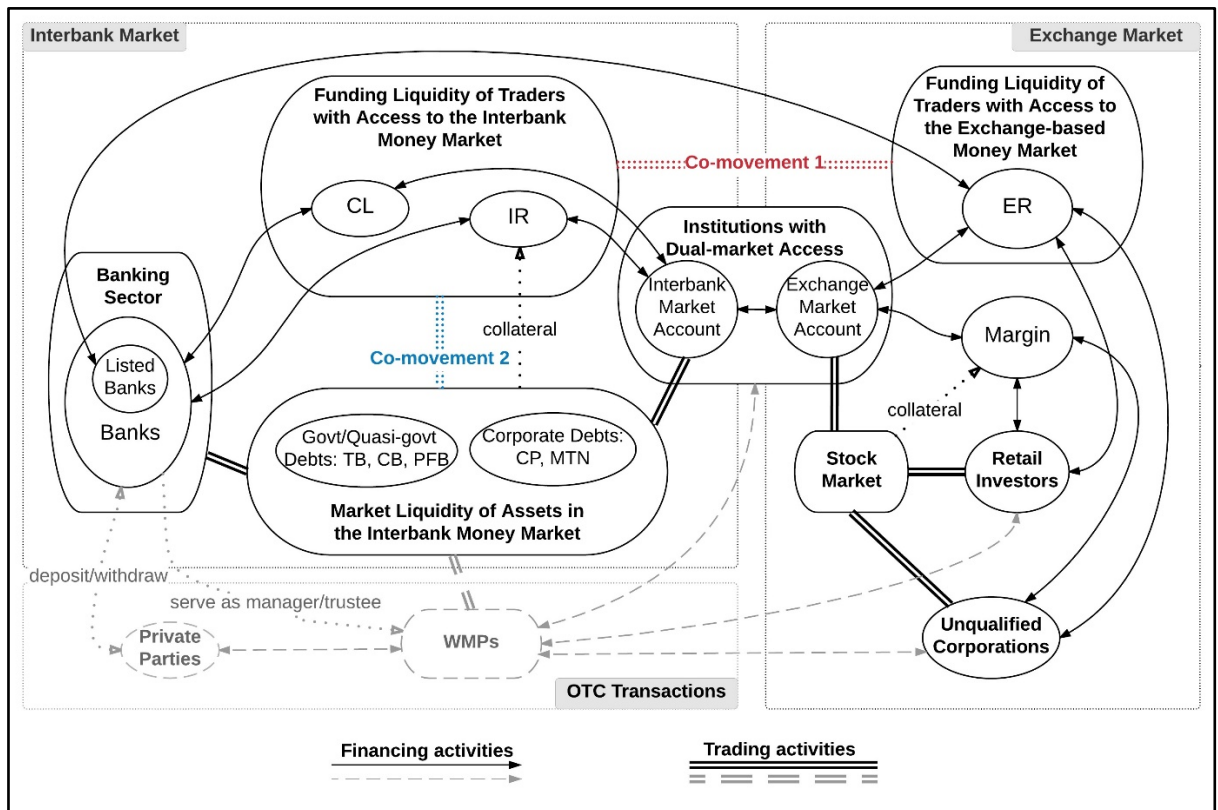

Figure 1. China's segmented money market and how it connects different traders and assets. (Source: Authors)

Because the money market is a submarket within the overall credit market, the segmentation in China's money market is a direct result of the artificial division of China's credit market into two parts: the interbank credit market mainly reserved for banks and institutional investors, and the exchange-based credit market mainly reserved for investors with access to China's two stock exchanges, which was previously described by Fan and Zhang (2007). Although the interconnectivity between the two markets has been greatly improved over the years, several major restrictions remain in place. With the exception of the 16 publicly listed commercial banks, no other banks are allowed to trade in the exchange market. Even for these 16 banks, the access to the exchange market has been partial, because they were prohibited from conducting exchange-based repo (ER) transactions until July, 2014 and are still prohibited from trading stocks and convertible bonds. On the other hand, the interbank credit market is generally off-limits to retail investors. Currently, only non-bank institutions, such as insurance companies, securities firms, fund management companies and qualified corporations can have relatively unrestricted access to the two markets. These entities are allowed have accounts in both markets, can trade securities (including stocks) and borrow funds in both markets, and can transfer funds and certain debt securities between their accounts in the two markets, albeit with cumbersome procedures and delays. Under the current market setting, the banks can still supply funds to the stock market by financing non-bank institutional investors in the interbank and exchange-based money markets, despite the market segmentation.

Along with the differentiated market access for different participant types, the security types in the two markets are also differentiated. Almost all short-term debt securities in the realm of money market are exclusively traded in the interbank money market, except for the short-term treasury bonds (TBs) that are occasionally issued in the exchange market. The major security types in the interbank money market include short-term TBs, short-term policy financial bonds (PFBs), which is a popular type of quasi-government bonds issued by government-backed policy banks, short-term central bank bills (CBs) used by the People's Bank of China (PBOC) as an open market operation tool, and short-term corporate debts in the forms of commercial papers (CPs) and medium-term notes (MTNs). Because these securities cannot be traded in or transferred into the exchange market, where stocks are traded, they cannot be held in the same portfolio with stocks. As a result, the direct interaction between their market liquidity and the market liquidity of stocks are significantly limited.

In addition, the funding sources for participants in the two money markets are also differentiated. There are two main funding sources in the interbank money market. Most of the institutions with access to this market can borrow with interbank repos (IRs) collateralized by interbank market assets. Qualified institutions can also obtain short-term funds from the interbank call loan (CL) market, an uncollateralized market mainly created for banks to trade their cash reserves with each other in order to satisfy the reserve requirement set by the PBOC, similar to the function of the Federal Funds market. The CL market also shares some commonalities with the Eurodollar market, because its benchmark interest rate Shibor is determined by a board of banks like the Libor, and not targeted by the central bank. In practice, non-bank institutions and smaller regional banks may find it difficult to borrow from the CL market due to its high credit requirement, thus having to rely more on the IR market for funding needs. This is evident by Cassola and Porter (2013)'s finding that small banks are usually net borrowers in the IR market but net lenders in the CL market, while the opposite is true for large national banks. In the exchange-based money market, traders can finance with ERs collateralized by credit assets in the exchange market. However, this funding source comes with significant restrictions, particularly for the retail investors. First, repo collaterals are difficult to accumulate in the exchange market, because exchange-based bonds are largely held by major financial institutions. Secondly, there are stringent requirements for retail investors to borrow with repo, such as having at least 500, 000 yuan in net asset value and a track record of active bond trading.

Given all the restrictions in China's public financial markets, the over-the-counter (OTC) "wealth-management products"[5] (WMPs) become an important supplement to them, because the regulators do not prohibit certain financial innovations by the banks through WMPs. For example, a bank can create a WMP funded by private parties, including many of its depositors, to finance the margins provided by securities firms to their customers. The off-balance-sheet status of these WMPs allows the banks to supply extra funds to the exchange-based credit and stock markets, skirting the reserve and capital requirements imposed on their balance-sheets and the restrictions in the public markets. Due to the non-standard nature of WMPs and their limited data availability, we do not explicitly account for their impact in this study. However, the transmission of liquidity shocks via WMPs could be partially captured by our analysis based on data from the two public money markets, because the WMPs are partially funded by the banking sector, and some of traders financed by WMPs might have access to the public money markets at the same time.

---

[5] A term used in China to represent uninsured financial products sold by banks and other financial institutions.

The Co-movements 1 and 2 highlighted in Figure 1 are the main focus in the rest of this paper. Co-movement 1 represents the co-movement between the funding liquidity of traders funded by the interbank money market and the funding liquidity of traders funded by the exchange-based money market. It can be viewed as a proxy for the interaction between the funding liquidity conditions of the banking sector and the stock traders, which is the primary transmission path for liquidity shocks between the banking sector and the stock market as we have discussed. We expect Co-movement 1 to strengthen when a liquidity shock affects either side of the market segmentation and drives up cross-market fund movements. Co-movement 2 represents the co-movement between the funding liquidity of traders funded by the interbank money market and the market liquidity of short-term assets traded and used as repo collateral in the interbank money market. Although Co-movement 2 is not a major transmission path for shocks between the banking sector and the stock market, its strength can serve as a key indicator for the stress level at the banking sector. Because the short-term debt securities in the interbank market do not have direct interaction with stocks, their market liquidity is expected to be mainly affected by the funding liquidity of their traders, who are most likely the banks being confined in the interbank market and having no access to the other short-term assets. Adding the fact that short-term debt securities are the easiest to liquidate in a credit market, we expect Co-movement 2 to rise sharply when the funding of the banks are tightened to a point that they have to sell a substantial amount assets to deal with cash shortage. In summary, strengthening Co-movement 1 is an indication that the funding liquidity of the banking sector is more closely tied to the funding liquidity of stock traders, while strengthening Co-movement 2 is a confirmation that a liquidity shock capable of affecting the stability of the banking sector has taken place. If Co-movements 1 and 2 strengthen at the same time, it is a strong evidence that either a liquidity shock originated from the stock market is being transmitted into the banking sector, or a liquidity shock originated from the banking sector is being transmitted into the stock market.

## 3. Methodology

In order to examine the two types of liquidity/illiquidity co-movements described in Section 2, we need to specify empirical measures for the funding liquidity of traders in China's interbank money market and exchange-based money market, as well as the market liquidity of assets in the interbank money market. We then need to construct a model to capture the co-movements among these liquidity measures. In the next three subsections, we detail our liquidity measures, our data and our empirical model.

### 3.1. Measuring funding liquidity and market liquidity

For traders in the interbank money market, we use two variants of liquidity measures to represent their funding liquidity, one based on IR and one based on CL. The measure based on IR is more representative of the funding liquidity of small regional banks and non-bank institutions in the interbank money market, because they are more likely to be able to borrow in the IR market than the CL market as mentioned in Section 2.2. The measure based on CL is more representative of the funding liquidity of large national banks. For traders in the exchange-based money market, we use the liquidity measure of ER to represent their funding liquidity. We also include two variants of liquidity measures of assets in the interbank money market, one based on PFB and one based on CP. The measure based on PFB is more representative of the market liquidity of short-term government debts, while the measure based on CP is more representative of the market liquidity of short-term corporate debts.

We specify our liquidity measures in a way that is consistent with Brunnermeier and Pederson (2009)'s theoretical model and similar measures in the U.S. market. In Brunnermeier and Pederson's (2009) model, the illiquidity level of a security is represented by the absolute deviation of its market price from its fundamental value. Because the prices of debt securities are usually quoted in interest rates rather than prices, it is more convenient for us to measure the price deviations for debt securities based on interest rate spreads. An example of such rate spread is the Libor-OIS spread, which is used in numerous studies on the 2007 U.S. subprime crisis as a proxy for risk and liquidity condition in the money market (see Frank, Gonzalez-Hermosillo and Hesse (2008), Brunnermeier (2009) and Gorton and Metrick (2012)). The Libor-OIS spread is the difference between the U.S. Dollar Libor rate, a reference interest rate for unsecured interbank lending of Eurodollar, and the fixed rate for Overnight Indexed Swap (OIS) that tracks the Fed Funds effective rate. It can be considered the premium of a bank's borrowing cost in the money market over a baseline rate subjected to little risk and liquidity concerns. It seems straightforward for us to adopt a Chinese equivalent to the Libor-OIS spread, i.e. the difference between Shibor and the fixed rate of an overnight interest rate swap (IRS) similar to the OIS, as the measure of illiquidity for CLs in China. However, there are two potential issues with this approach. First, the roles of Libor-OIS spread and its Chinese equivalent as liquidity measures can be complicated by the impact of credit risk on Libor and Shibor, due to their nature as unsecured rates. Both Hesse and Frank (2009) and Smith (2012) have shown that Libor-OIS spread contains a component for liquidity risk (i.e. the risk of solvent counterparties having unexpected cash constraints to meet obligations) as well as a component for credit risk (i.e. the risk of loss due to insolvency of counterparties). Yet, because major banks in China are perceived to be backed by the government against insolvency due to their major or partial government ownership, the credit risk component in a Shibor-based spread is likely trivial. Secondly, the availability of China's IRS rate data is rather spotty due to infrequent IRS transactions in China, making the Shibor-OIS spread impractical. To circumvent this issue, we use the spread between Shibor and the TB yield instead. The Shibor-TB spread can be considered the Chinese equivalent to the TED spread, i.e. the difference between Libor and the U.S. Treasury bill (T-bill) yield, which is commonly used analogously to the Libor-OIS spread.

Similarly, we use the IR-TB spread, ER-TB spread, CP-TB spread and PFB-TB spread to represent the illiquidity levels of IRs, ERs, CPs and PFBs, respectively. Most of these rate spreads have their counterparts in studies on the U.S. market. The IR-TB and ER-TB spreads are comparable to the U.S. Repo-Tbill spread suggested by Bai, Krishnamurthy and Weymuller (2016) as a preferred measure for the liquidity condition in the U.S. banking sector. The CP-TB spread in China is an imperfect equivalent to the U.S. ABCP-Tbill spread used by Frank, Gonzalez-Hermosillo and Hesse (2008) as a measure of liquidity condition in the U.S. ABCP market. Unlike the ABCPs in the U.S., the CPs in China are all non-asset-backed variants, thus are not protected against credit risk. To minimize the impact of credit risk, we specifically use the yield of AAA-rated CP to calculate the CP-TB spread. The PFB-TB spread is unique to the Chinese market. The PFBs are quasi-government bonds issued by the government-designated policy banks to support large public projects, such as infrastructure projects, while the TBs are directly issued by the government. Unlike certain short-term TBs that can be traded in both markets, the short-term PFBs are only traded in the interbank market, and are generally priced with a premium over TBs. According to Wan (2006), the difference in pricing between PFB and TB is not only attributed to their liquidity difference but also attributed to the difference in tax policies for their interest returns. To remedy this issue, we explicitly account for the impact of tax differences in our model.

**3.2. The data**

The five rate spreads specified in Section 3.1 are calculated from the weekly averages of six daily interest rates with 1-month maturity for the period between June 17, 2011, the earliest date when all the six rates are available, and April 8, 2016. Our original data consists of 1-month Shibor published by NIBFC, the weighted average 1-month interbank repo rate and the 1-month CP, PFB and TB yields collected from CEIC Data, and the daily closing rate of the standardized exchange repo GC028 listed in Shanghai Stock Exchange. We specifically choose the 1-month maturity to maximize data availability, because repo transactions with maturities over a month are infrequent in China, while the PFBs and TBs with less than a month maturities are nonexistent. We also use weekly averages to eliminate excess noise in the daily data, which may be caused by the day of week liquidity effect and the routine open market operations by the PBOC. The six interest rates and the calculated 1-month rate spreads are displayed in Figure 2 and Figure 3 respectively.

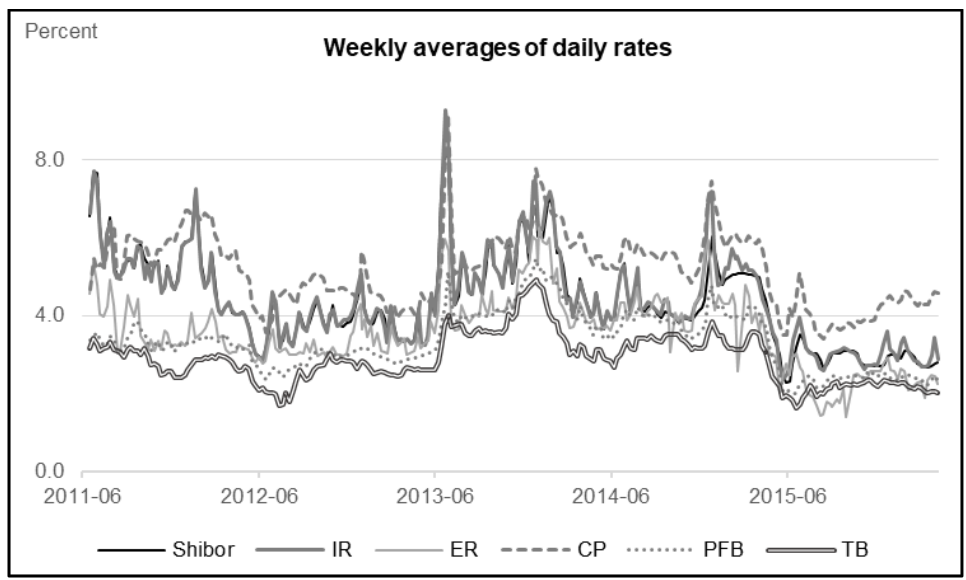

Figure 2. Weekly averages of the six daily interest rates. (Source: CEIC Data, NIBFC, Sohu Finance)

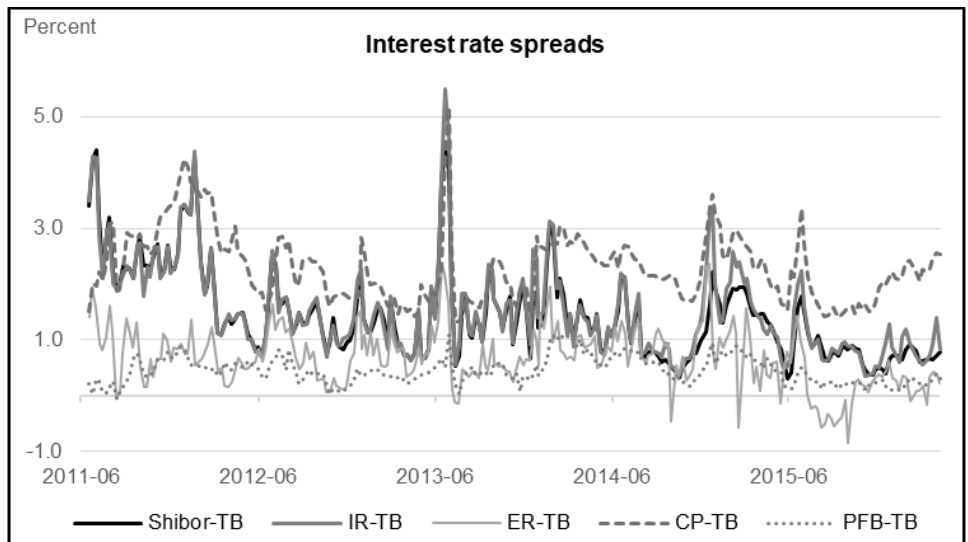

Figure 3. The five interest rate spreads as liquidity measures. (Source: Authors)

According to Figure 2, the TB yield is generally the lowest among the six interest rates while the CP yield is generally the highest, which is consistent with their risk and liquidity characteristics. Meanwhile, the difference between Shibor and IR rates is very small, suggesting the credit risk component in Shibor is indeed trivial. In Figure 3, our liquidity measures (i.e. the illiquidity levels represented by the rate spreads) fluctuate violently with a range up to 500 basis points, which is unusual among similar rate spreads in the U.S.

### 3.3. Modelling the co-movements of liquidity measures

Because we adopt two variants of funding liquidity measures as well as two variants of market liquidity measures for the interbank money market, our model is tasked to measure six variants of liquidity co-movements rather than two, including two versions of Co-movement 1 and four versions of Co-movement 2. The six variants of liquidity co-movements are visualized in Figure 4.

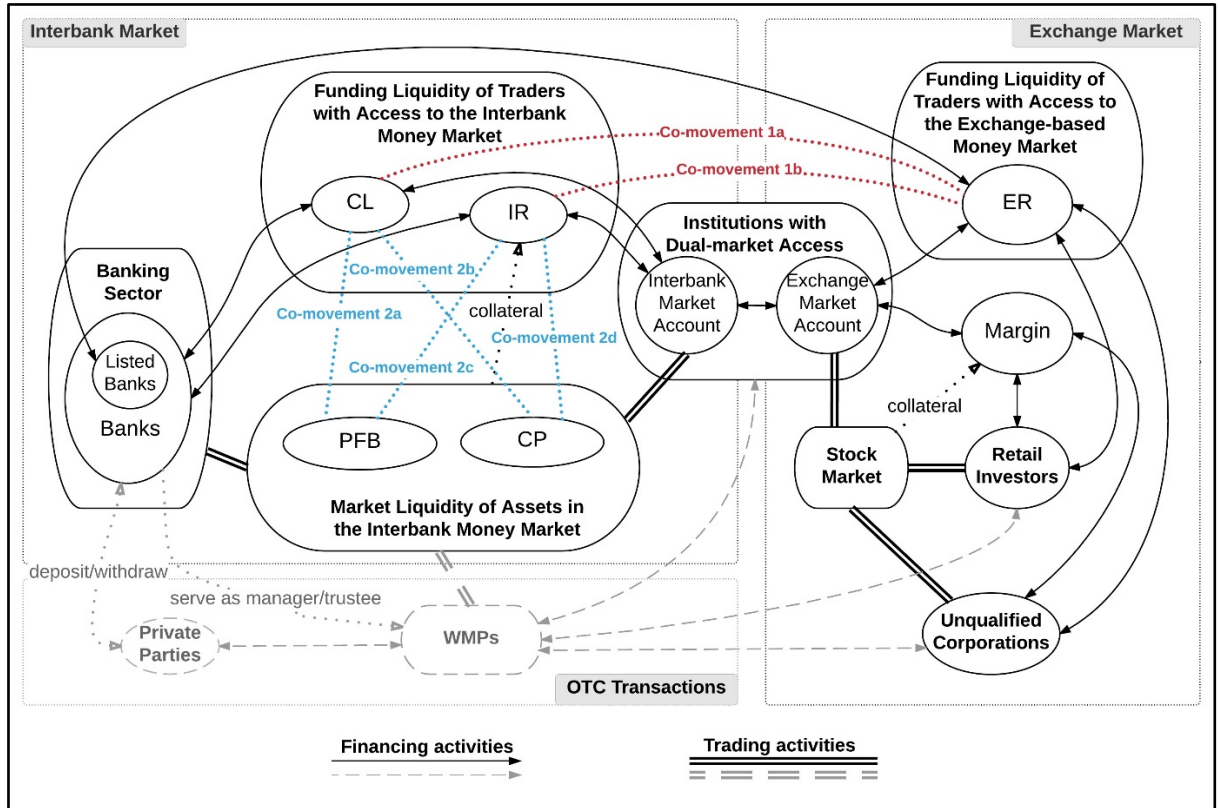

Figure 4. The six liquidity co-movements to be measured by the empirical model. (Source: Authors)

In Brunnermeier and Pederson (2009)'s theoretical model, the co-movement between the liquidity conditions of two securities is represented by the covariance of the liquidity measures of these two securities. Based on this concept, the co-movements among the five liquidity measures in this study are represented by their covariance matrix or correlation matrix, which can be viewed as the covariance matrix adjusted for individual volatilities. Obviously, the six co-movements highlighted in Figure 4 are elements in the correlation matrix. By adopting this concept, our task then becomes estimating the time-varying correlation matrix among the five liquidity measures. A multivariate volatility model, such as the Dynamic Conditional Correlation (DCC) model specified by Engle (2002), seems to be ideal for the task. In fact, a modified DCC-GARCH model has already been used by Frank, Gonzales-Hermosillo and Hesse (2008) to empirically examine the spillovers of liquidity shocks during the 2007 subprime crisis. The DCC-GARCH model and similar time-varying correlation multivariate GARCH models have also been used to study the other aspects of cross-market and cross-asset co-movements, such as the co-movements in returns and volatilities. For example, Chiang, Joen and Li (2007) study the cross-market financial contagion among Asian markets based on a DCC-GARCH model of stock returns. Yang, Zhou and Wang (2009) document the time-varying correlation between stock and bond market returns in the U.S. for the past 150 years using a smooth transition multivariate GARCH model. Arouri, Jouini and Nguyen (2011) examine the volatility transmission between oil and stock markets in Europe and the U.S. using a VAR-GARCH model.

Most of these studies use multiple bivariate models to estimate the time-varying correlation for each pair of assets or markets. Instead of following their approach, we estimate the time-varying correlation matrix of the five liquidity measures with a single penta-variate DCC-GARCH model. We expect the six co-movements to interact with each other, so we estimate them together in one model. In our model, the conditional means of the five rate spreads are assumed to follow an AR(1) process and the conditional covariance matrix is assumed to follow a multivariate DCC(1, 1)-GARCH(1, 1) process. The AR(1) model for the PFB-TB spread include an additional variable to account for the impact of different tax policies for PFB and TB. The mean equations for the interest rate spreads are specified as

$$x_{i,t} = \mu_i + \phi_i x_{i,t-1} + r_{i,t}, \quad i = 1, 2, 3, 4 \tag{1}$$

$$x_{5,t} = \mu_5 + \phi_5 x_{5,t-1} + \tau B_t + r_{5,t} \tag{2}$$

$$r_{i,t} = \sqrt{h_{i,t}}\, \varepsilon_{i,t}, \quad i = 1, 2, 3, 4, 5 \tag{3}$$

where $x_{1,t}$ is the Shibor-TB spread, $x_{2,t}$ is the IR-TB spread, $x_{3,t}$ is the ER-TB spread, $x_{4,t}$ is the CP-TB spread, $x_{5,t}$ is the PFB-TB spread, $B_t$ is the 1-month TB yield, $h_{i,t}$ is the conditional variance of innovations, and $\varepsilon_{i,t}$ is a standardized disturbance with mean zero and variance one. The intuition behind using the TB yield variable to capture the impact of tax difference is straightforward. Suppose PFB has comparable risk and liquidity characteristics as TB, but its interest payment is taxable as opposed to the tax-free TB, the theoretical interest rate for PFB $I_{5,t}$ should satisfy the following relationship $I_{5,t} = 1/(1-T) * B_t$ where $T$ is the tax rate. In this case, $x_{5,t} = I_{5,t} - B_t = T/(1-T) * B_t$. Thus, the difference between $I_{5,t}$ and $B_t$ contributed by tax difference should be proportional to the yield of TB. The reason for not using a specific tax rate $T$ to directly calculate $T/(1-T)$ is because tax policies regarding interest payments from PFBs differ for different type of institutions. In this case, estimating $T/(1-T)$ with a parameter $\tau$ is a preferable approach to approximate the average impact of tax across different types of institutions.

Our DCC(1, 1)-GARCH(1, 1) model for the conditional correlations is specified based on Engle (2002)'s formulation. But unlike the original model, we assume that the innovation vector $\boldsymbol{r_t} = [r_{1,t}, r_{2,t}, r_{3,t}, r_{4,t}, r_{5,t}]'$ follows a multivariate Student's t distribution instead of a multivariate normal distribution to account for fat tails. Suppose we define the covariance of the innovation as $\boldsymbol{H_t} = \boldsymbol{D_t R_t D_t}$ where $\boldsymbol{D_t} = diag\{\sqrt{h_{i,t}}\}$ and $\boldsymbol{R_t}$ is the time varying correlation matrix. Our DCC model can be specified as

$$\boldsymbol{r_t}|_{t-1} \sim t(0, \boldsymbol{D_t R_t D_t}) \tag{4}$$

$$\boldsymbol{D_t^2} = diag\{\omega_i\} + diag\{\kappa_i\} \circ \boldsymbol{r_{t-1} r_{t-1}'} + diag\{\lambda_i\} \circ \boldsymbol{D_{t-1}^2}, i = 1,2,3,4,5 \tag{5}$$

$$\boldsymbol{\varepsilon_t} = \boldsymbol{D_t^{-1} r_t} \tag{6}$$

$$\boldsymbol{Q_t} = (1 - \alpha - \beta)S + \alpha \boldsymbol{\varepsilon_{t-1} \varepsilon_{t-1}'} + \beta \boldsymbol{Q_{t-1}} \tag{7}$$

$$\boldsymbol{R_t} = diag\{\boldsymbol{Q_t}\}^{-1/2} \boldsymbol{Q_t} diag\{\boldsymbol{Q_t}\}^{-1/2} \tag{8}$$

$$S = E[\boldsymbol{\varepsilon_t \varepsilon_t'}] \tag{9}$$

where $\boldsymbol{\varepsilon_t} = [\varepsilon_{i,t}]$, $\boldsymbol{Q_t} = [q_{i,j,t}]$, $\boldsymbol{R_t} = [\rho_{i,j,t}]$, $\{\omega_i, \kappa_i, \lambda_i, \alpha, \beta\}$ are the parameters to be estimated and ∘ denotes Hadamard product. Equation (5) governs the GARCH process and Equation (7) governs the DCC process. Equation (7) can be interpreted as a GARCH(1,1) process for the conditional correlation matrix. Each element of $\boldsymbol{Q_t}$ satisfies $q_{i,j,t} = \bar{\rho}_{i,j} + \alpha(\varepsilon_{i,t-1}\varepsilon_{j,t-1} - \bar{\rho}_{i,j}) + \beta(q_{i,j,t-1} - \bar{\rho}_{i,j})$ with $\bar{\rho}_{i,j}$ being the unconditional correlation between $\varepsilon_{i,t}$ and $\varepsilon_{j,t}$.

Based on the assumption of Student's t distribution in Equation (4), we construct the likelihood function of the whole AR-DCC-GARCH system specified in Equations (1)-(9), and jointly estimate the AR parameters $\{\mu_i, \phi_i, \tau\}$, the DCC-GARCH parameters $\{\omega_i, \kappa_i, \lambda_i, \alpha, \beta\}$ and the shape parameters of the Student's t distribution with the Quasi-Maximum Likelihood (QML) procedure specified in Engle (2002).

## 4. Empirical results

In this section, we present the results from our model, explore the six variants of liquidity co-movements represented by the estimated conditional correlations, and discuss what these co-movements tell us about the transmission of liquidity shocks into and out of China's banking sector in the last five years, particularly during the two market events mentioned in Section 1.

### 4.1. Model fitting and parameter estimates

Before our model is estimated by the QML procedure, the five interest rate spreads are tested for stationarity by the augmented Dickey-Fuller test and the Phillips-Perron test. Neither tests support the existence of unit roots in the five series, so the levels of the rate spreads are directly used to fit the model. The QML results are tabulated in Table 1.

|  | **Shibor-TB** | **IR-TB** | **ER-TB** | **CP-TB** | **PFB-TB** |
|---|---|---|---|---|---|
| | Conditional Mean: AR(1) | | | | |
| $\mu_i$ | 0.914*** (0.132) | 1.052*** (0.133) | 0.636*** (0.090) | 1.764*** (0.464) | 0.947*** (0.347) |
| $\phi_i$ | 0.888*** (0.033) | 0.816*** (0.039) | 0.735*** (0.055) | 0.948*** (0.038) | 0.944*** (0.027) |
| $\tau$ | | | | | -0.224* (0.128) |
| | Conditional Variance: GARCH(1, 1) | | | | |
| $\omega_i$ | 0.006 (0.005) | 0.024** (0.011) | 0.016 (0.017) | 0.029** (0.013) | 0.002 (0.002) |
| $\kappa_i$ | 0.432*** (0.098) | 0.526*** (0.157) | 0.125 (0.084) | 0.756** (0.354) | 0.316** (0.143) |
| $\lambda_i$ | 0.567*** (0.094) | 0.473*** (0.084) | 0.768*** (0.171) | 0.243*** (0.079) | 0.575*** (0.155) |
| | Conditional Correlation: DCC(1, 1) | | | | |
| $\alpha$ | 0.033** (0.017) | | | | |
| $\beta$ | 0.945*** (0.038) | | | | |

Table 1. Parameter estimates of the penta-variate AR(1)-DCC(1,1)-GARCH(1,1) system.
*: significant at 10% level; **: significant at 5% level; ***: significant at 1% level. (Source: Authors)

The empirical behaviors of the five rate spreads are well summarized by our model because most of the parameters are statistically significant at 5% level, including the serial correlation in conditional means, the volatility clustering in conditional variances and the fluctuations in conditional correlations. However, the negative parameter for the tax factor in the PFB-TB spread is puzzling. According to Section 3.3, the difference between the yields of PFB and TB contributed by the tax factor should be proportional to the TB yield. Given a positive tax rate, when the yield of TB increased, the tax related component in the PFB-TB spread should increase rather than decrease as suggested by the negative $\tau$. One possible explanation for the negative $\tau$ is that higher yields in the credit market might have changed the composition of market participants in the credit market. According to Zhao (2015), the interest returns from PFBs are taxable for most financial institutions but tax-free for fund management companies. When yields in the credit market increased, fund management companies might invest more in credit assets including PFBs. As a result, tax-free investors might constitute a larger proportion of PFB holders when yields were high, thereby lowering the average tax rate for PFB investors as well as the required premium of the PFB yield over the TB yield to offset the difference in tax.

**4.2. The liquidity co-movements measured by conditional correlations**

The two versions of Co-movement 1 and four versions of Co-movement 2 represented by the estimated conditional correlations among our liquidity measures are visualized in Figures 5 and 6, respectively. Before presenting the details, it is worth noting that each date marked on the figures is either a Friday or the last trading day in a week if it is not Friday, due to the fact that we use weekly averages of daily rates to estimate our model.

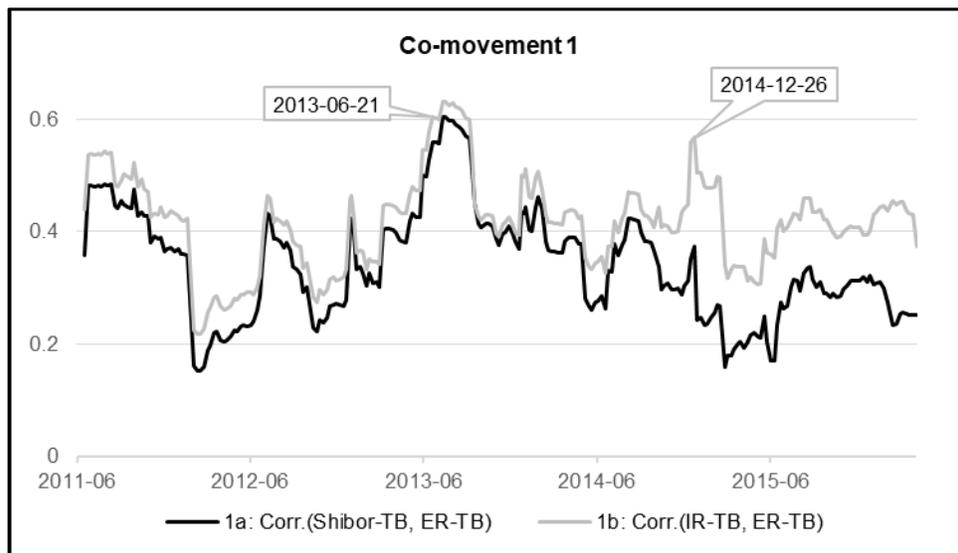

Figure 5. The co-movement of funding liquidity between the banking sector and stock traders. (Source: Authors)

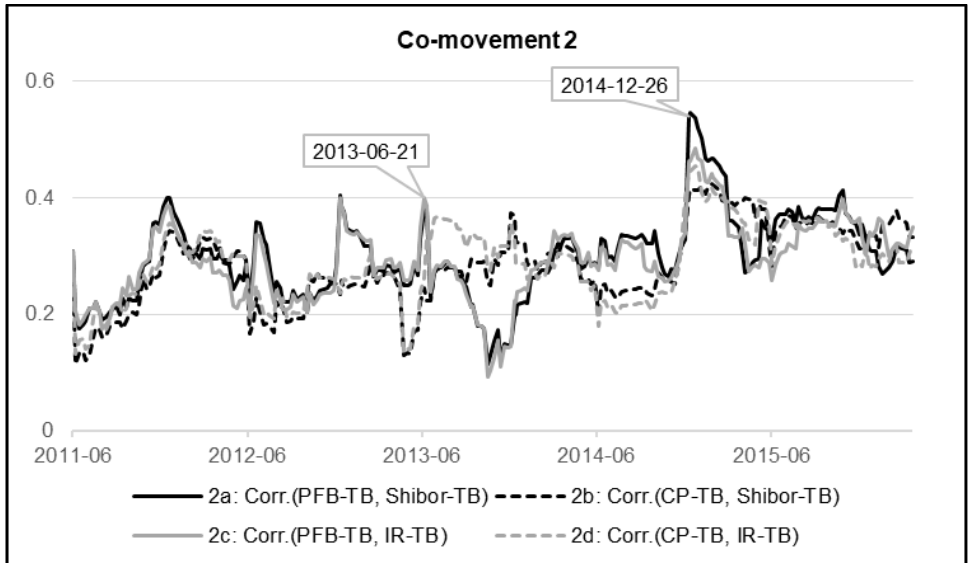

Figure 6. The co-movement between the funding liquidity of the banking sector and the market liquidity of major short-term debt securities in the interbank money market. (Source: Authors)

The two sets of empirical correlations estimated by our DCC model are all positive over the whole sample period, ranging from around 0.1 at the low points to over 0.6 at the high points. These fluctuating positive correlations suggest that the co-movements among our liquidity measures are substantial, and their strength varies much during the sample period, which is consistent with what we have expected. Meanwhile, from Figures 5 and 6, we notice a few distinct features[6] in the patterns of the liquidity co-movements in close proximity to the timing of the two market events in focus. The most prominent feature in Co-movement 1 (i.e. the cross-market funding liquidity co-movement) is the peak of its strength at around mid-2013, which is close to the timing of the "Shibor Shock" that took place in the week ending on June 21, 2013. There is a second peak in the strength of Co-movement 1 based on IR (Co-movement 1b in Figure 5) around the end of 2014, which is in the middle of a period of drastic stock market movements. But the same peak is much less noticeable in Co-movement 1 based on CL (Co-movement 1a in Figure 5), and the gap between the two versions of Co-movement 1 seems to have widen significantly shortly before the peak and remain wider than it used to be ever since. The most prominent feature in Co-movement 2 (i.e. the co-movement of funding liquidity and market liquidity in the interbank market) is the spike in its strength around the end of 2014, which is captured by all the four versions of its empirical measures (see Figure 6). The timing of the spike coincides with the second peak in Co-movement 1b that we have suspected to be related to the stock market event. There is also a minor spike in the strength of Co-movement 2 around the time of the "Shibor Shock", but it is less standout compared to the first peak in Co-movement 1. In addition, the versions of Co-movement 2 based on PFB (Co-movement 2a and 2c in Figure 6) and based on CP (Co-movements 2b and 2d in Figure 6) alternate in strength before and after the "Shibor Shock". These patterns and features have important implications on the transmission of liquidity shocks during the two market events in focus, which are discussed in Section 4.3.

---

[6] The distinct features in the liquidity co-movements discussed in Sections 4.2 and 4.3 are robust for different starting dates of our sample. For example, even if we estimate our model based on a shortened data set from 2012 onward and recreate Figures 5 and 6, we will still see the same distinct features in the new figures.

**4.3. The transmission of liquidity shocks during the two market events**

By examining the liquidity co-movements captured by our empirical model in conjunction with the regulatory actions and policy changes surrounding the two market events (i.e. the rollercoaster ride in China's stock market between late-2014 and mid-2015 and the "Shibor Shock" in June, 2013), we are able to better understand how the liquidity shocks related to the two market events were transmitted between China's banking sector and stock market via money market transactions. Our findings strongly suggest that liquidity shocks can be transmitted both ways between the banking sector and the stock market via the money market, and the transmission of shocks can be affected by various decisions of China's policy makers. Our narratives on the two events are presented separately in Section 4.3.1 and Section 4.3.2.

**4.3.1. The rollercoaster ride of China's stock market in 2014-2015**

The dramatic movements in China's stock market between late-2014 and mid-2015 are illustrated by the CSI 300 Index in Figure 7. The two types of liquidity co-movements corresponding to the period of Figure 7 are shown in Figures 8 and 9. The CSI 300 Index a major stock index in China that tracks the performance of 300 stocks listed in Shanghai Stock Exchange and Shenzhen Stock Exchange. In this study, we specifically focus on the period between October, 2014, when the CSI 300 index rose above 2500 points for the first time since May, 2013, and June, 2015, when the index eventually reached above 5300 points but quickly collapsed back to the 3000-4000-point range. The patterns shown in Figures 7, 8 and 9 clearly indicate that there was a stock-market-related liquidity shock affecting the banking sector via the money market in late-2014. Surprisingly, however, this shock seems to be related to the heating-up of the stock market rather than the collapse of the stock market.

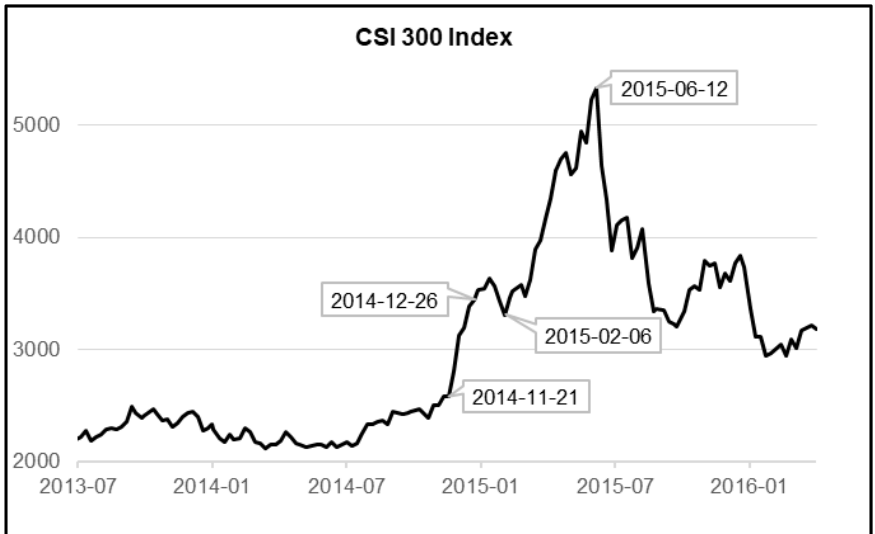

Figure 7. CSI 300 Index between July 2013 and April 2016. (Source: CEIC Data)

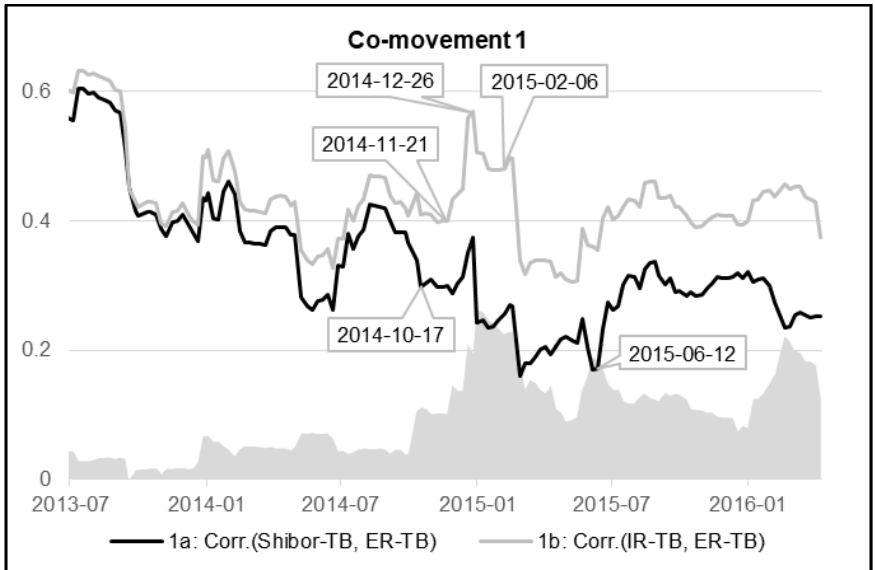

Figure 8. Co-movement 1 between July 2013 and April 2016. Shadow shows the different between the two versions. (Source: Authors)

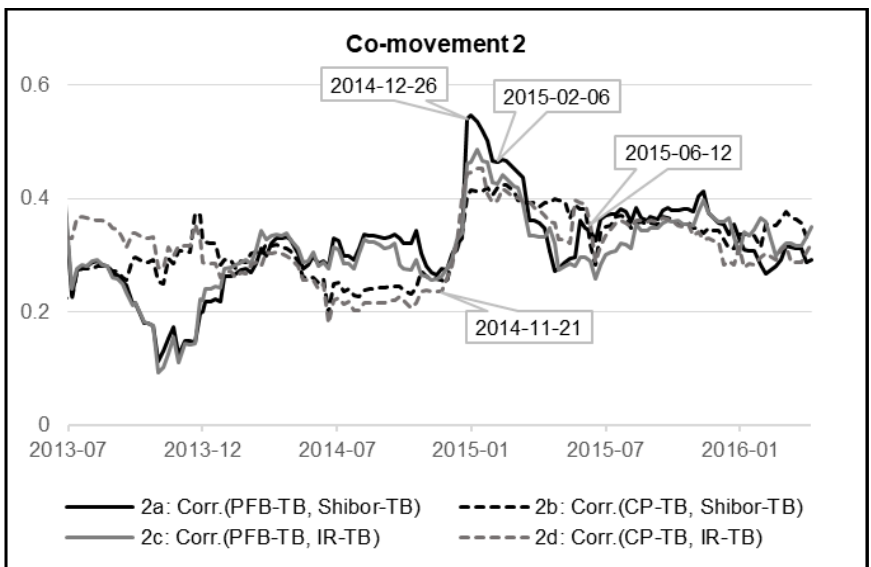

Figure 9. Co-movement 2 between July 2013 and April 2016. (Source: Authors)

As we have discussed at the end of Section 2.2, strengthening Co-movement 1 is a sign that the funding liquidity of traders in the interbank market is more closely tied to the funding liquidity of traders in the stock exchanges; strengthening Co-movement 2 is a sign that the funding liquidity in the banking sector is drying up; and the two co-movements strengthening at the same time is a double confirmation that there is a liquidity shock being transmitted between the banking sector and the exchange market. From Figures 7, 8 and 9, we indeed observe that, between the week ended on November 21, 2014 and the end of 2014, both Co-movement 1 and Co-movement 2 strengthened sharply while the CSI 300 Index skyrocketed from around 2600 points to around 3400 points, which is a strong indication that a liquidity shock related to the stock market had affected the funding liquidity of the banking sector through the money market during that period. It is unlikely that the shock being transmitted was originated from a problem in the stock market, because otherwise the stock market would not have gone up so quickly at the same time. Instead, we conjecture that it was due to the funds in the exchange market being all tied-up in equity assets in the white-hot stock market, drying up the funds available for the further financing of stock traders. The funding illiquidity in the exchange market was transmitted into the interbank market when traders with dual-market access started bringing funds from the interbank market to the exchange market in response to the growing funding constraint in the exchange market, driving up Co-movement 1. As the funding condition in the interbank market tightened up, the banking sector might have been forced to recoup cash by liquidating some assets in interbank market, subsequently driving up Co-movement 2.

The patterns of the two co-movements and their correspondence with the timing of two major policy changes in October and November, 2014, respectively, suggest that the policy makers might have played an important role in the initiation and transmission of the liquidity shock, because they made it easier for stock traders to access funds in the banking sector and provided incentive for them to borrow more. As mentioned in Section 4.2, the gap between the versions of Co-movement 1 based on CL (1a) and based on IR (1b) widens notably since October, 2014 (see Figure 8). The most likely contributor of this shift is a new rule issued by the PBOC on October 17, 2014[7], which allowed qualified non-financial corporations with at least 30 million yuan in net assets to access the interbank credit market.

---

[7] http://www.pbc.gov.cn/zhengcehuobisi/125207/125227/125963/2810732/index.html (in Chinese)

This policy change lowered the barriers between the two money markets and provided new funding opportunities for many commercial entities that previously only had access to funds in the exchange market. Given that these newcomers to the interbank market were more likely to borrow from the IR market rather than the unsecured CL market mainly reserved for high-credit financial institutions, the cross-market flows of funds carried by them should mainly take place between the two repo markets. Thus, we expect the cross-market liquidity co-movement based on IR to strengthen relative to the cross-market liquidity co-movement based on CL following the new rule, which is indeed what we see in Figure 8. Meanwhile, the turning point of Co-movements 1 and 2 at around November 21, 2014 is also not a coincidence. On November 22, 2014, the PBOC lowered the benchmark interest rates in China for the first time since 2012[8], signaling major monetary easing and providing a strong incentive for more borrowing. Following the rate change, the strength of Co-movements 1 and 2 immediately jumped up, reflecting tightening funding liquidity conditions in both money markets, while the upward movement in the CSI 300 Index started to accelerate. This is a substantial evidence that funds were leaving the banking sector via the interbank money market, being brought into the exchange market by entities with dual-market access, and being invested in the stock market. The much stronger reaction in Co-movement 1b compared to Co-movement 1a (see Figure 8) is again an indication that the liquidity drain from the banking sector mainly took place in the IR market. In summary, our empirical results suggest that the policy change in October widened the money-market-based pathway for funds to flow between the banking sector and the exchange market. The rate change in November then triggered a major outflow of funds from the banking sector to the stock market via money market transactions.

The timing of the alleviation and dissipation of the liquidity stress, as implied by the sharp drops in the strength of the two liquidity co-movements at the end of 2014 and

---

[8] http://www.pbc.gov.cn/zhengcehuobisi/125207/125213/125440/125835/2892213/index.html (in Chinese)

mid-2015 (see Figures 8 and 9), is also consistent with two policy decisions that aimed to release liquidity into the banking sector. During the week ending on December 26, 2014, the PBOC injected more than 300 billion yuan into the banking sector via Short-term Liquidity Operations (SLOs)[9], which was a temporary relief for the funding illiquidity in the banking sector. On February 4, 2014, the PBOC lowered the reserve requirement ratio (RRR) for all banks for 0.5%, which was the first RRR adjustment since May, 2012[10] and a major relaxation of the funding constraint faced by the banking sector. This decision meant that an amount equal to 0.5% of all the renminbi deposits in China were freed up from the banking sector's mandatory reserves held by the central bank, which was a huge amount. The RRR change was followed by a sharp drop in the strength of both Co-movements 1 and 2, as well as a new round of rally in the stock market after a brief pull-back in January, suggesting that the liquidity conditions were quickly improving over the board.

The above narrative demonstrates that the two types of liquidity co-movements estimated by our model are capable of capturing the transmission and dissipation of liquidity shocks between China's banking sector and stock market via its segmented money market. During the rapid downfall of the stock market after its peak on June 12, 2015 (see Figure 7), the liquidity co-movements displayed in Figures 8 and 9 again showed signs of illiquidity being transmitted between the two money markets and into the banking sector. Co-movement 1 strengthened immediately after June 12, 2014 and Co-movement 2 followed suit a week later. However, the reactions of Co-movements 1 and 2, particularly Co-movement 2, in the aftermath of the stock market collapse were much milder compared to the one-month period before the end of 2014, suggesting that the funding condition of the banking sector were not severely hampered by the violent downturn in the stock market. Based on this observation, we suspect that the retail investors restricted from borrowing in the two money markets and their financing activities via OTC funding channels (refer to Figure 4) subjected to much less regulatory restrictions might have played a major role in the finale of the stock market rally before it ended in mid-June. In this case, the margin calls triggered by the freefalling equity prices would mainly affect the funding liquidity of traders not reliant on funds from the two money markets and their creditors not directly financed by the banking sector, thus diverting the transmission of liquidity shock away from the two public money markets and limiting how much shock could be captured by Co-movements 1 and 2. Our conjecture is supported by a report from China Securities Regulatory Commission (CSRC) [11] issued in 2015, which stated that the number of new accounts in China's stock market opened in the first quarter of 2015 grew by over 400% quarter-over-quarter, and over 90% of these new accounts had a net asset value less than 500 thousand yuan, the legal minimum for margin accounts set by the CSRC. This report confirms that most of the new investors entering the stock market in early 2015 were indeed small retail investors that did not even have indirect access to funds in the two money markets via margins provided by their brokers, and would have to resort to OTC funding sources if they wanted to add leverage to their portfolios. Therefore, it is important to understand and monitor the transmission of liquidity shocks via OTC transactions outside of the two public money markets in order to prevent unexpected illiquidity contagion not reflected by the fluctuations in the public markets. A future study on this specific subject is needed once more data on China's OTC market becomes available.

---

[9] http://www.pbc.gov.cn/zhengcehuobisi/125207/125213/125431/125478/2810439/index.html (in Chinese)
[10] http://www.pbc.gov.cn/zhengcehuobisi/125207/125213/125434/125798/2875174/index.html (in Chinese)

**4.3.2. The "Shibor Shock"**

The "Shibor Shock" was a short-lived cash crunch in China's interbank money market that took place around mid-June in 2013 and reached its climax on June 20, 2013, when the Shibor rates shot up to record high levels. For example, the overnight Shibor jumped from 7.66% on June 19 to 13.44% on June 20, an unimaginable level for an overnight interbank

---

[11] http://www.csrc.gov.cn/pub/newsite/tzzbh1/tbtzzjy/tbfxff/201504/t20150428_275686.html (in Chinese)

rate. There was no official account for the event, but the media attributed it to the shortage of funds in the banking sector since early June and the PBOC temporarily withholding its liquidity provision to banks aiming to curb excess credit expansion (The Economist 2013a, 2013b). If this was indeed the case, then the liquidity shock was clearly originated from the banking sector, and would be reflected in strengthening Co-movement 2 (i.e. the co-movement between the funding liquidity and market liquidity in the interbank market) around the time of the "Shibor Shock". Meanwhile, if this liquidity shock was transmitted from the banking sector to exchange-based traders by money market transactions and eventually affected the liquidity in the stock market, Co-movement 1 (i.e. the cross-market funding liquidity co-movement) would also strengthen, and the stock market performance would be negatively affected around the time of the shock. All these features are confirmed by the patterns of the liquidity co-movements estimated by our model between May, 2013 and July, 2013 and the CSI 300 Index during the same period, which are displayed in Figures 10-13. Therefore, our empirical results provide strong evidences that a liquidity shock originated from the banking sector can be transmitted into the stock market via money market transactions. By further examining patterns of the liquidity co-movements surrounding the event, we are able to not only provide a narrative on the spreading of the "Shibor Shock", but also help verify some of the claims made by the media regarding what might have caused the liquidity shortage in the first place and how the banking sector and policy makers responded to the "Shibor Shock".

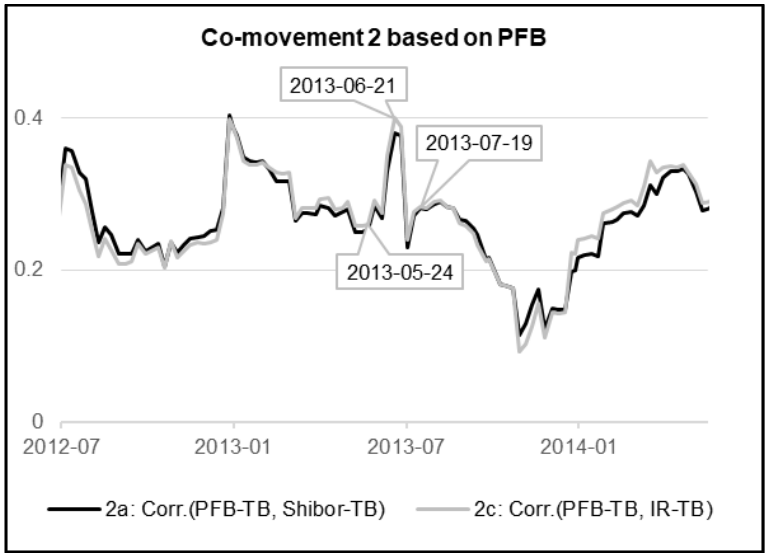

Figure 10. Co-movement 2 based on PFB between July 2012 and May 2014. (Source: Authors)

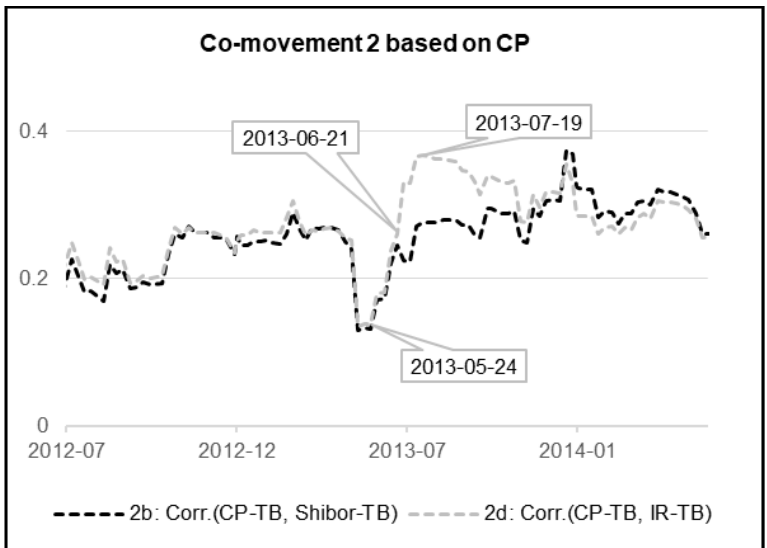

Figure 11. Co-movement 2 based on CP between July 2012 and May 2014. (Source: Authors)

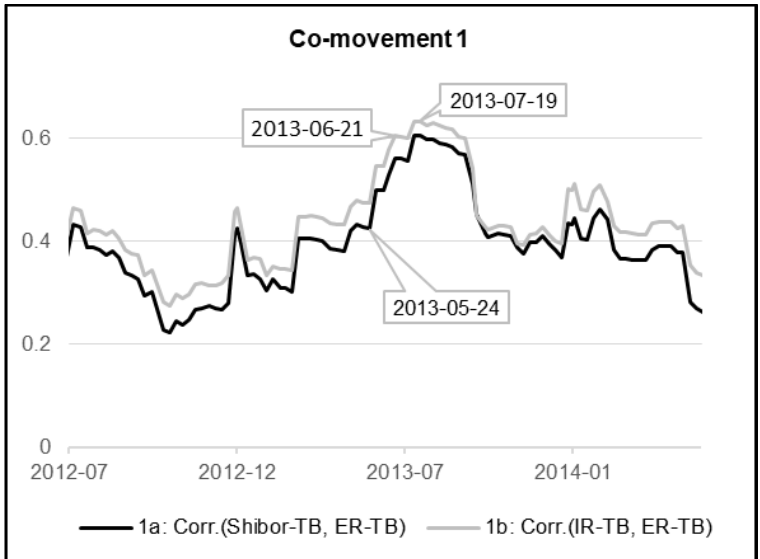
Figure 12. Co-movement 1 between July 2012 and May 2014. (Source: Authors)

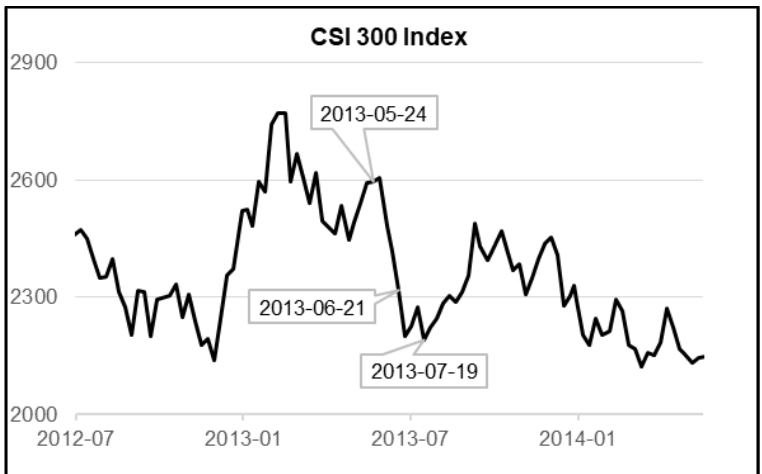
Figure 13. CSI 300 Index between July 2012 and May 2014. (Source: CEIC Data)

The liquidity co-movements shown in Figures 10-12 seem to suggest that the funding condition in the banking sector started tightening rapidly since late May. Media reports speculated that the PBOC actually knew about the situation and reacted to it, but its reaction was the opposite of what the banks expected. According to a leaked statement from a private PBOC meeting, the PBOC was alarmed by an apparent surge in lending in the first ten days in June, when the banks in China added almost 1 trillion yuan to their loan books, more than the amount they typically lend in a whole month; and the PBOC concluded that some banks were expecting a fresh government stimulus and had "positioned themselves in advance" (The Economist, 2013b). The same article commented that the PBOC might have misread the banks' intentions, because 70% of the 1 trillion yuan in new loans were short-term discounted bills intended to facilitate transactions between commercial enterprises, which were likely smuggled off balance-sheets at smaller banks and only resurfaced in early June after the regulator tightened up accounting at these banks. However, our results suggest that the PBOC's opinion had its own merit. Even if the new loans were mainly discounted bills, it appeared that the banks did stretched their balance-sheets to finance these bills, which happened as early as in May. Our liquidity measure for CP is a good proxy for the liquidity of discounted bills, because they are both short-term corporate loans serving similar purposes. We separate the versions of Co-movement 2 based on PFB (i.e. Co-movements 2a and 2c) and the versions based on CP (i.e. Co-movement 2b and 2d) into two figures to show the discrepancies between them before the "Shibor Shock" and afterwards. By only looking at the CP versions Co-movement 2 (see Figure 11), we may think that the funding constraint for the banking sector were suddenly relaxed in early May, 2013, because the strength of Co-movements 2b and 2d dropped to the lowest level since late 2011 at that time (refer to Figure 6). Yet, when we look at the PFB versions of Co-movement 2 for the same period (see Figure 10), we cannot find obvious clues for the major improvement of the funding condition in the banking sector. This could happen if the banks continued to make it easier for companies to borrow short-term funds from them (i.e. improving liquidity of CPs) despite their funding liquidity were not improved (i.e. the liquidity of IRs and CLs not improved) and they did not want to buy more low-interest PFBs (i.e. the liquidity of PFBs not improved), hence the much weakened Co-movements 2b and 2d but relatively stable Co-movements 2a and 2c in May, 2013.

The liquidity co-movements also reveal that the transmission of the liquidity shock originated from the banking sector continued beyond the "Shibor Shock" and lasted well into July, 2013. The commonalities and differences in the patterns of different types/versions of the liquidity co-movements during and after the "Shibor Shock" provide strong indications that the banks changed their preferences and behaviors in response to the "Shibor Shock". All four versions of Co-movement 2 strengthened sharply during the 2-3 weeks before June 21, 2013, but their movements started to differ significantly since the "Shibor Shock" (see Figures 10-12). If we only look at the Co-movement 2 based on PFB (see Figure 10), we are inclined to think that the "Shibor Shock" quickly dissipated and merely had a transient impact. However, the co-movements shown in Figures 11 and 12 seem to tell a different story, in which the illiquidity in the banking sector continued to intensify and being transmitted into the exchange market until late July. This apparent discrepancy can be explained by a phenomenon called "flight to quality", a term coined by Brunnermeier and Pederson (2009). In our case, flight to quality means that, when funding becomes scarce for the banks, they will favor safer and more liquid assets and will prioritize their cutback on assets with higher volatility. It also means that the market liquidity of assets with higher volatility will be more sensitive to changes in the funding liquidity of the banks, i.e. the versions of Co-movement 2 based on assets with higher volatility would be stronger than the versions of Co-movement 2 based on assets with lower volatility. Although the "Shibor Shock" was temporarily eased when the PBOC provided liquidity support for selected banks by June 25, 2013[12], the liquidity pressure was still high in the banking sector, because there was no monetary stimulus in sight from the PBOC and the banks could no longer retain the mindset that PBOC would back them up whenever they were short of liquidity. This was the ideal setup for the flight to safety phenomenon to appear. Facing the persistent liquidity pressure, the banks preferred to hold more of their assets in PFBs rather than CPs, because the PFB was much less volatile than the CP (see Figure 2) and risk-free. In this case, even if they no longer needed to fire-sale their assets,

---

[12] http://www.pbc.gov.cn/goutongjiaoliu/113456/113469/2868130/index.html (in Chinese)

they might continue to sell their CP holdings while buying PFBs. Therefore, it is not surprising to see that the CP-based versions of Co-movement 2 (see Figure 11) had been stronger than the PFB-based versions of Co-movement 2 (see Figure 10) for close to six months after the "Shibor Shock". We also notice from Figure 11 that the difference in strength between Co-movement 2b (i.e. the liquidity co-movement between CP and CL, the major funding source for large banks) and Co-movement 2d (i.e. the liquidity co-movement between CP and IR, the major funding source for small banks) was substantially wider during the five months after the "Shibor Shock" than the rest of the sample period. We consider it an indication that the flight to safety phenomenon was more pronounced for small banks than large banks, possibly due to CPs and other short-term corporate loans having more weights in the assets of small banks than large banks prior to the "Shibor Shock". Similar to the versions of Co-movement 2 based on CP, the strength of Co-movement 1 continued to rise after the "Shibor Shock" until it reached its highest point in history in mid-July (see Figure 12). We suspect it was driven by the banks pulling back the funds they previously supplied to traders with dual-market access, including institutional investors in the stock market, in order to cope with liquidity pressure, because the CSI 300 Index had been dropping since early June until reaching its lowest point in 2013 in mid-July, some 15% lower than the beginning of June (see Figure 13). These fund movements can be attributed to the flight-to-safety phenomenon as well, because the banks may have preferred to hold more cash and safe assets such as PFBs and TBs in the interbank market than financing traders with exposure to the riskier and more volatile exchange market. In summary, by assessing the estimated liquidity co-movements before, during and after the "Shibor Shock", we are not only able to verify the transmission of liquidity shock from the banking sector to the stock market via money market transactions, but also able to provide important insights on the involvement of the policy makers in the build-up of this event and the reactions of the banking sector in the aftermath of this event.

## 5. Conclusion

This study is the first ever assessment on how the transactions in China's segmented money market may facilitate the transmission of liquidity shocks across different parts of China's financial system. In this study, we provide an anatomy on the money-market-based transmission paths for liquidity shocks, create an array of empirical measures to gauge the funding liquidity of the banking sector and the stock traders as well as the market liquidity of short-term credit assets, and develop a multivariate DCC-GARCH model to capture the transmission of liquidity shocks in action during the last five years. The results from our empirical model provide strong evidences that money market transactions can indeed transmit liquidity shocks back and forth between China's banking sector and the stock market that it does not have direct access to, despite the regulatory restrictions imposed on the banks and the segmentation existed in China's money market. The evidences enable us to better understand how liquidity shocks were transmitted across different parts of China's financial system during the rollercoaster ride of China's stock market between late-2014 and mid-2015 and the "Shibor Shock" in mid-2013. They also provide us important insights on the crucial roles played by China's banking sector and policy makers during these two events.

What we have found in this study can be regarded as both a testimony for the improved interconnectivity between China's interbank market and exchange market, and a warning sign for the potential risk associated with it. It can serve as a strong reminder for China's policy makers and market participants that money market transactions are capable of allowing a liquidity shock to circumvent various regulatory restrictions and segmentations in China's financial system and result in a widespread illiquidity contagion, thus presenting a major threat to the overall stability of China's financial system. Therefore, it is important for the policy makers and regulators in China to closely monitor how funds are flowing through the two money markets in China to prevent a system-wide liquidity crisis like the one that took place in the U.S. in 2007. It is also important for the entities that actively engage in money market transactions in China, particularly the banking sector, to prepare for the unexpected illiquidity contagion associated with trading and financing in the money market.